\documentclass[twocolumn,epjc3]{svjour3}  
\smartqed  
\RequirePackage{graphicx}
\RequirePackage{fix-cm}
\RequirePackage{amssymb}
\RequirePackage{amsmath}
\RequirePackage{float}
\RequirePackage{xcolor}
\RequirePackage[numbers,sort&compress]{natbib}
\RequirePackage[colorlinks,citecolor=blue,urlcolor=blue,linkcolor=blue]{hyperref}
\usepackage[final]{microtype}
\journalname{Eur. Phys. J. C}
\begin{document}

\title{Big Bang Nucleosynthesis constraints on $f(T,L_m)$ gravity}


\author{Daniel F.P. Cruz\thanksref{e1,addr1,addr2}
        \and David S. Pereira\thanksref{e2,addr1,addr2}
        \and Francisco S.N. Lobo\thanksref{e3,addr1,addr2}
        \and Jos\'e P. Mimoso\thanksref{e4,addr1,addr2}
}


\thankstext{e1}{e-mail: fc56684@alunos.ciencias.ulisboa.pt}
\thankstext{e2}{e-mail: djpereira@ciencias.ulisboa.pt (corresponding author)}
\thankstext{e3}{e-mail: fslobo@ciencias.ulisboa.pt}
\thankstext{e4}{e-mail: jpmimoso@ciencias.ulisboa.pt}

\institute{Departamento de Física, Faculdade de Ciências da Universidade de Lisboa, Edifício C8, Campo Grande, 1749-016 Lisboa, Portugal\label{addr1}
           \and
           Instituto de Astrofísica e Ciências do Espaço, Edifício C8, Campo Grande, 1749-016 Lisboa, Portugal\label{addr2}
}

\date{Received: date / Accepted: date}

\maketitle

\begin{abstract}
We investigate Big Bang Nucleosynthesis (BBN) in the framework of $f(T,L_m)$ gravity, where the gravitational Lagrangian depends on the torsion scalar $T$ and the matter Lagrangian $L_m$. Working within a semi-analytical BBN strategy, we encode departures from GR through the expansion-rate ratio $Z\equiv H/H_{\rm GR}$ evaluated at a characteristic freeze-out temperature and combine this with the freeze-out condition and the observationally inferred abundances of deuterium and helium-4 to constrain the free parameters of three representative EFT-motivated $f(T,L_m)$ models. A distinctive aspect of $f(T,L_m)$ cosmology is that the explicit $L_m$ dependence can induce an effective energy exchange between the standard component and the modified-gravity sector; we therefore derive the corresponding interaction term $Q$ and restrict our analysis to the adiabatic regime $\varepsilon\equiv |Q_{\rm rad}/(4H\rho)|\ll 1$ throughout the BBN window, ensuring internal consistency of the temperature-based BBN mapping. Finally, to connect the radiation-era constraints with the late-time background, we present a two-fluid (dust+radiation) analysis showing how the $L_m$-dependent corrections decouple as $\ell\propto (1+z)^4\to 0$, yielding torsion-only ($f(T)$ or TEGR) cosmologies at late times on the GR-connected branch. Our results provide transparent first-pass BBN bounds on torsion--matter EFT corrections and identify viable parameter regions consistent with early-Universe data providing a controlled starting point for further early-Universe phenomenology in $f(T,L_m)$ gravity.
\end{abstract}

\section{Introduction}
Big Bang Nucleosynthesis (BBN) refers to the epoch in the early Universe during which light nuclei were synthesized from the primordial plasma of protons and neutrons. This process occurred within the first few minutes of cosmic evolution, approximately during the interval $t \sim 1\,$s to $t \sim 10^3\,$s after the Big Bang. In particular, weak freeze-out occurs at temperatures of order $\mathcal{T}\sim 1\,{\rm MeV}$, corresponding to cosmic times of order seconds, while efficient light-nuclei formation begins later, around $t\sim 10^2\,{\rm s}$, when the temperature has fallen to $\mathcal{T}\sim 0.1\,{\rm MeV}$. At such temperatures the density of free nucleons was sufficiently high to enable nuclear reactions, while the decreasing photon density allowed these nuclei to survive photodissociation. Consequently, the first stable light nuclei were formed, leading to the primordial production of hydrogen ($^1\!H$), deuterium ($D$), helium-4 ($^4\!He$), and lithium-7 ($^7\!Li$).

Within the framework of the standard cosmological model, Standard BBN (SBBN) provides remarkably successful predictions for the primordial abundances of helium-4 and deuterium, both of which are in excellent agreement with observational data over a wide range of astrophysical environments. However, a persistent discrepancy remains in the case of lithium-7: the predicted primordial abundance of $^7\!Li$ is a factor of about three higher than what is inferred from spectroscopic observations of old, metal-poor stars in the Galactic halo. This well-known tension, commonly referred to as the ``lithium problem'' \cite{Fields:2011zzb,Miranda:2025wcp,Makki:2024sjq,ParticleDataGroup:2024cfk}, continues to motivate theoretical investigations. Proposed resolutions range from astrophysical explanations, such as stellar depletion processes, to more exotic scenarios involving new physics beyond the Standard Model of particle physics or modifications of gravitational dynamics. For a comprehensive discussion of the physics of BBN, its observational constraints, and the status of the lithium problem, see \cite{Grohs:2023voo, Cooke:2024nqz, Steigman:2012ve}.
{In this work we adopt a semi-analytical BBN strategy in which the impact of modified cosmological dynamics is encoded through the expansion-rate ratio $Z\equiv H/H_{\rm GR}$ evaluated at a characteristic freeze-out temperature. In practice we take $T_f=0.55\,{\rm MeV}$ as a representative value and translate departures from GR into parameter bounds using the same philosophy as recent semi-analytical analyses of BBN in modified gravity, see e.g.~\cite{Ge:2024tsx}. This approach provides a transparent and analytically tractable first-pass constraint and is particularly useful as a
model-selection tool before undertaking a full reaction-network likelihood analysis.}

Indeed, since BBN provides one of the earliest and most sensitive probes of the expansion history of the Universe, it serves as a stringent test for alternative cosmological models beyond General Relativity (GR). In particular, many recent works have focused on the interplay between teleparallel gravity extensions and the predictions of primordial element abundances. For instance, constraints on $f(T)$ models, where $T$ denotes the torsion scalar, were systematically studied in \cite{Capozziello:2017bxm}. Extensions of teleparallel gravity have also been investigated in this context. Theories of the form $f(T,\hat{T})$ \cite{Harko:2014aja}, where $\hat{T}$ represents the trace of the energy–momentum tensor, introduce direct matter–torsion couplings and thereby enrich the underlying phenomenology \cite{Mishra:2023onl}. Similarly, $f(T,T_G)$ models \cite{Asimakis:2022kfk}, with $T_G$ denoting the teleparallel equivalent of the Gauss–Bonnet invariant, extend the theory by incorporating higher-order torsional contributions. While BBN offers a stringent early-time probe of 
$f(T)$ gravity, independent cosmological studies have also examined its implications for background evolution, perturbations, and late-time cosmic acceleration \cite{Bahamonde:2021gfp}.

Parallel developments in curvature-based modified gravity have also been tested against BBN. Notably, scalaron fields arising in $f(R)$ gravity theories \cite{Talukdar:2024qny} and quadratic energy–momentum squared gravity \cite{Jang:2024jso} have been explored. Further constraints have been derived for symmetric teleparallel gravity extensions. In $f(\mathbf{Q})$ gravity \cite{Anagnostopoulos:2022gej}, where $\mathbf{Q}$ denotes the non-metricity scalar, the modified Friedmann equations directly impact the nucleosynthesis era. Theories such as $f(\mathbf{Q},T)$ gravity \cite{Bhattacharjee:2021hwm}, which incorporate explicit matter–geometry couplings, have been analyzed under BBN predictions, with observational consistency requiring specific parameter ranges. More recently, Weyl-type generalizations of $f(\mathcal{Q},T)$ gravity \cite{Ge:2024tsx} have been tested, again demonstrating that BBN remains a powerful diagnostic tool capable of constraining free parameters across a wide variety of modified gravitational frameworks.

{In this paper, we investigate the implications of BBN within the framework of $f(T,{L}_m)$ gravity. Analogous to its curvature counterpart \cite{Harko:2010mv}, the gravitational Lagrangian is generalized to a function of the torsion scalar $T$ and the matter Lagrangian density ${L}_m$, introducing additional matter--geometry couplings and extra degrees of freedom \cite{torsion-matter} that can modify the early-Universe expansion. We study representative EFT-motivated forms of $f(T,{L}_m)$ gravity and quantify their impact on the thermal and nuclear processes responsible for primordial element formation. Our main constraints are obtained by requiring consistency with the observationally inferred deuterium and ${}^4$He abundances, which are widely regarded as the most robust probes of the BBN expansion history. Finally, since the non-minimal dependence on $L_m$ can induce an effective energy exchange between the standard component and the modified-gravity sector, we complement the BBN bounds with a robustness check based on the interaction term $Q$ in the radiation continuity equation. In the parameter regions of interest we verify that $\left|Q/(4H\rho)\right|\ll 1$, ensuring that the radiation-dominated interpretation underpinning the semi-analytical BBN mapping remains self-consistent within $f(T,L_m)$ gravity.}

A natural question is whether BBN-based bounds on $f(T,L_m)$ parameters also determine the viability of the theory for inflation or for late-time cosmic acceleration. In this work we do not pursue a global fit across all epochs. Rather BBN constitutes a particularly clean \emph{early-Universe consistency test}: any viable torsion--matter operator must (i) reproduce an expansion rate compatible with the light-element abundances in the MeV era and (ii) preserve the standard radiation-dominated thermal history to a good approximation, which we explicitly monitor through the interaction measure $\varepsilon\equiv |Q_{\rm rad}/(4H\rho)|\ll1$. Once these necessary early-time viability conditions are enforced, the resulting parameter ranges can be used as physically motivated priors for other high-energy mechanisms within the same framework such as inflation, reheating, baryogenesis. In particular, torsion--matter couplings in $f(T,L_m)$ gravity can drive additional out-of-equilibrium effects relevant to the origin of the matter--antimatter asymmetry, such as gravitational baryogenesis, while remaining compatible with BBN constraints \cite{CRUZ2026117304}. Thus, beyond constraining the MeV-era expansion history, the bounds and methodology derived here provide a controlled starting point for further early-Universe phenomenology in $f(T,L_m)$ gravity.

A related recent study considered BBN and late-time acceleration in $f(T,L_m)$ gravity for the particular inverse-torsion model $f(T,L_m)=\alpha T_0^2/T+\beta L_m$, using the BBN freeze-out-temperature bound together with late-time observational data sets~\cite{SwagatMishra:2025scq}. The present work is complementary, but its physical scope is different. A key ingredient of our analysis is the choice $L_m=P$. With this choice, the explicit matter-Lagrangian sector is pressure-activated: it is relevant during the radiation era, where $P=\rho/3$, but it dynamically decouples in the late-time dust regime, where $P_m\simeq0$. Consequently, in our framework there is no genuinely late-time $f(T,L_m)$ matter-coupling effect in the pressureless matter era, apart from the negligibly small radiation pressure; the late-time limit is instead TEGR or a torsion-only $f(T)$ cosmology, depending on the benchmark model. This is why our late-time discussion is not
intended as a global low-redshift data fit of a torsion--matter coupling, but rather as a consistency check showing how the explicit $L_m$ corrections decouple after the radiation era.

Our main aim is therefore to use BBN as a clean early-Universe probe of torsion--matter EFT corrections. Unlike single-model early--late background
reconstructions, we study three representative classes of operators: a separable torsion plus matter-sector correction, a genuinely mixed torsion--matter monomial, and a logarithmic torsion sector supplemented by a screened matter correction. Moreover, we do not rely only on the freeze-out-temperature constraint; we also impose the deuterium and ${}^4{\rm He}$ bounds on the expansion-rate ratio $Z\equiv H/H_{\rm GR}$. Finally, because the explicit $L_m$ dependence can induce an effective energy exchange between the radiation bath and the modified-gravity sector, we derive the corresponding interaction term $Q_{\rm rad}$ and impose the adiabaticity condition $\varepsilon\equiv |Q_{\rm rad}/(4H\rho)|\ll1$. This additional consistency requirement ensures that the standard temperature-based BBN mapping used in the analysis remains valid.

This work is organized as follows: The action and field equations of $f(T,L_m)$ gravity are introduced in section~\ref{f(T)_gravity}. In section~\ref{Standart_BBN}, we present the standard BBN formalism, explore the constraint for modified gravity models using the observable abundance of light elements, and in section~\ref{sec:BBN_FTL} we apply the constraints for three specific $f(T,L_m)$ models. Finally, we discuss and conclude our work in section~\ref{Conclusion}.

\section{$f(T,L_m)$ Gravity}\label{f(T)_gravity}

In teleparallel gravity, the fundamental dynamical variables are the tetrads $e^{\ \mu}_{A}$, which constitute an orthonormal basis for the tangent space at every spacetime point. Greek indices $(\mu, \nu, \ldots)$ are used to denote spacetime coordinates, whereas Latin indices $(A, B, \ldots)$ label tangent-space (local Lorentz) coordinates, with both sets of indices running from $0$ to $3$. The spacetime metric can then be constructed from the tetrads as
\begin{equation}
    g_{\mu\nu}=\eta_{AB}e^{A}_{\ \mu} e^{B}_{\ \nu}\, ,
\end{equation}
where $\eta_{AB} = {\rm diag}(1, -1, -1, -1)$ is the metric tensor of
Minkowski spacetime in Cartesian coordinates.

Instead of the Levi-Civita connection, teleparallel gravity employs the curvature-free Weitzenböck connection \cite{Pereira}, defined in terms of the tetrads as $\Gamma^{\lambda}_{\ \mu\nu} \equiv e_A^{\ \lambda}\,\partial_\nu e^{\ A}_{\mu}$. Using this connection, the torsion tensor is defined as the antisymmetric part of the connection in its lower indices
\begin{equation}
    T^\rho_{\mu\nu}\equiv\Gamma^\rho_{\mu\nu}-\Gamma^\rho_{\nu\mu}=e_A^{\ \rho}(\partial_\mu e^A_{\ \nu}-\partial_\nu e^A_{\ \mu})\, .
    \label{torsion}
\end{equation}

Analogous to the Ricci scalar $R$, the Lagrangian of the teleparallel theory of gravity is the torsion scalar $T$, given by
\begin{equation}
    T\equiv T^\rho_{\ \mu\nu}S_\rho^{\ \mu\nu}=\frac{1}{4}T^{\rho\mu\nu}T_{\rho\mu\nu}+\frac{1}{2}T^{\rho\mu\nu}T_{\nu\mu\rho}-T_{\rho\mu}^{\ \ \ \rho}T^{\nu\mu}_{\ \ \ \nu}\,,
    \label{Torsion Scalar}
\end{equation}
where $S_\rho^{\ \mu\nu}$ is the superpotential and is defined as
\begin{equation}
    S_\rho^{\ \mu\nu}=\frac{1}{2}\left(K^{\mu\nu}_{\ \ \ \rho}+\delta^\mu_\rho T^{\alpha\nu}_{\ \ \ \alpha}-\delta^\nu_\rho T^{\alpha\mu}_{\ \ \ \alpha} \right)\,.
\end{equation}
The cotorsion tensor $K^{\mu\nu}_{\ \ \ \rho}$ is provided by
\begin{equation}
    K^{\mu\nu}_{\ \ \ \rho}=-\frac{1}{2}(T^{\mu\nu}_{\ \ \ \rho}-T^{\nu\mu}_{\ \ \ \rho}-T_\rho^{\ \mu\nu})\,.
\end{equation}

{In this work we extend the teleparallel equivalent of GR (TEGR) by promoting the torsion scalar in the gravitational sector to
\begin{equation}
T \;\longrightarrow\; T+f(T,L_m)\,,
\end{equation}
so that the action becomes
\begin{equation}
S=\frac{1}{16\pi G}\int d^4x\, e\Big[T+f(T,L_m)\Big]+\int d^4x\, e\,L_m\,,
\label{action}
\end{equation}
where $e\equiv \det(e^{A}{}_{\mu})=\sqrt{-g}$, $G$ is Newton's constant, and $M_{\rm Pl}\equiv (8\pi G)^{-1/2}\simeq 2.4\times 10^{18}\,{\rm GeV}$ denotes the reduced Planck mass. The matter Lagrangian $L_m$ is assumed to depend on the tetrads $e^{A}{}_{\mu}$ (and matter fields) but not on derivatives of the tetrads, which is the standard assumption in this class of teleparallel matter-coupled models.}

{A useful way to interpret the extra term $f(T,L_m)$ is as an effective-field-theory (EFT) correction to TEGR: in the same spirit in which GR can be regarded as a low-energy EFT~\cite{Donoghue:1994dn}, additional higher-dimensional operators built from the available low-energy scalars are expected to appear, with coefficients suppressed by an underlying heavy scale~\cite{Li:2018ixg}. Such operators (and, in particular, non-minimal couplings between the gravitational and matter sectors) may naturally be generated once heavy degrees of freedom are integrated out, and their impact can become more pronounced in high-energy regimes, e.g. at large energy densities or large characteristic torsion/expansion scales relevant to the early Universe. In a torsional formulation these operators can be organized in an EFT expansion involving the torsion scalar (and, more generally, the torsion tensor and its derivatives), and the EFT viewpoint for teleparallel gravity has been developed in detail in the literature~\cite{Li:2018ixg, Mylova:2022ljr,Harko:2014sja}. Within this perspective, the function $f(T,L_m)$ parametrizes the leading corrections and possible non-minimal torsion--matter couplings that can arise after integrating out heavy degrees of freedom, and it provides a compact phenomenological description of such effects.
}

From the variation of the action \eqref{action} with respect to the tetrads we get the field equations:
\begin{equation}
    \begin{split}
        &\left[e^{-1}\partial_\mu(ee_A^{\ \rho}S_\rho^{\ \mu\nu})-e_A^{\ \lambda}T^{\rho}_{\ \mu\lambda}S_\rho^{\ \nu\mu}\right](1+F_T)\\
        &+e_A^{\ \rho}S_\rho^{\ \mu\nu}(F_{TT}\partial_\mu T+F_{TL}\partial_\mu L_m)+e_A^{\ \nu}\left(\frac{T+f}{4}\right)\\
        &-\frac{1}{4}F_L\left(e_A^{\ \rho}\overset{em}{T}{}_\rho^{\ \nu}+e_A^{\ \nu}L_m\right)=4\pi G e_A^{\ \rho}\overset{em}{T}{}_\rho^{\ \nu}\,,
    \end{split}
    \label{field equations}
\end{equation}
 where $\overset{em}{T}{}_\rho^{\ \nu}$ is the energy-momentum tensor of a perfect fluid, defined as (see chapter 9.4 of \cite{Pereira})
\begin{equation} 
 \frac{\delta eLm}{\delta e^A_{\ \nu}}=-ee_A^{\ \rho}T_\rho^{\ \nu}
 \end{equation} 
 and $F_T=\frac{\partial f}{\partial T}$, $F_{TT}=\frac{\partial^2 f}{\partial T^2}$, $F_{TL}=\frac{\partial^2 f}{\partial T\partial L}$, $F_L=\frac{\partial f}{\partial L}$.

We consider a spatially flat FLRW Universe described by the metric
\begin{equation}
	ds^2 = dt^2 - a^2(t)\,\delta_{ij}\,dx^i dx^j,
\end{equation}
where $a(t)$ denotes the scale factor. This metric can be obtained from the diagonal tetrad
\begin{equation}
	e_A^{\ \mu} = \mathrm{diag}(1,\,a,\,a,\,a).
	\label{tetrad}
\end{equation}

Inserting the tetrad (\ref{tetrad}) into the field equations (\ref{field equations}), we obtain the following modified Friedmann equations
\begin{equation}
    H^2=\frac{8\pi G}{3}\rho-2H^2F_T+\frac{1}{6}F_L\left(\rho+L_m\right)-\frac{f}{6}\, ,
    \label{first Friedmann}
\end{equation}
\begin{equation}
   \begin{split}
        \Dot{H}=-4\pi G(\rho+P)-\Dot{H}(F_T-12H^2F_{TT})\\
        -HF_{TL}\partial_tL_m-\frac{1}{2}F_L\left(P+L_m\right)\,,
   \end{split}
\end{equation}
where $H = \dot{a}/a$ is the Hubble parameter, with the overdot denoting differentiation with respect to cosmic time. The effective energy density and pressure are represented by $\rho$ and $P$, respectively. Using the tetrad (\ref{tetrad}) together with equations (\ref{torsion}) and (\ref{Torsion Scalar}), the torsion scalar for the FLRW metric can be computed as
\begin{equation}
    T=-6H^2\,.
    \label{T}
\end{equation}

We can rewrite the above modified Friedmann equations as
\begin{equation}
    H^2=\frac{8\pi G}{3}\left(\rho+\rho_{MG}\right)\label{Hubble}\, ,
\end{equation}
\begin{equation}
    2\dot{H}+3H^2=-8\pi G \left(P+P_{MG}\right)\label{Raychu}\, ,
\end{equation}
where
\begin{equation}
    \rho_{MG}=\frac{3}{8\pi G}\left[-2H^2F_T+\frac{1}{6}F_L\left(\rho+L_m\right)-\frac{f}{6}\right]\label{rho_DE}\, ,
\end{equation}
and
\begin{eqnarray}
        P_{MG}&=-\rho_{MG}+\frac{1}{8\pi G}\Big[2\Dot{H}(F_T-12H^2F_{TT})
        \nonumber \\
        &+2H\partial_t(L_m)F_{TL}+F_L(P+L_m)\Big]\, .\label{Constrain1}
\end{eqnarray}

 For the following $f(T,L_m)$ models, we adopt the perfect-fluid matter Lagrangian in the form $L_m=P=w\rho$,where $w$ is the equation-of-state (EoS) parameter, following a standard prescription in models with non-minimal matter--gravity couplings \cite{Bertolami:2007gv,Bertolami:2008ab}. In minimally coupled GR, different on-shell choices for a perfect-fluid Lagrangian (e.g.\ $L_m=P$ or $L_m=-\rho$) can lead to the same stress--energy tensor; however, once the gravitational action depends explicitly on $L_m$, the choice becomes physically relevant because it directly controls the strength and interpretation of the matter--gravity coupling \cite{Harko:2010mv,Bertolami:2008ab,Bertolami:2008zh}. Our motivation for $L_m=P$ is twofold. First, it is a widely used and internally consistent choice in the non-minimal coupling literature \cite{Bertolami:2007gv,Bertolami:2008ab}. Second, it implements a clear phenomenological selection principle aligned with the scope of this work: the explicit $L_m$-dependent corrections are ``pressure-activated'' and therefore most relevant in the high-pressure radiation epoch that governs BBN. An immediate implication is that, in the late-time dust regime ($P\to0$), the explicit $L_m$-dependent contributions dynamically decouple ($L_m=0$), so the background evolution reduces for the majority of the model to a torsion-only sector (TEGR or an $f(T)$ deformation, depending on the model), for which late-time cosmological constraints have been extensively studied \cite{f(T),f(T)_and_cosmology,Bahamonde:2021gfp}. In the subsection~\ref{subsec:latetime_limit} we elaborate on this point. Consequently, our BBN bounds should be interpreted as early-Universe constraints on torsion--matter corrections in the radiation era, rather than as a complete late-time cosmological fit. A global analysis combining the present BBN constraints with late-time data would require specifying and fitting the torsion-only sector and, if desired, exploring alternative fluid choices such as $L_m=-\rho$ (which keep the coupling active in the matter era); we leave such a comprehensive late-time study for future work.

{\subsection{Impact of $f(T,L_m)$ on energy transfer}}

Equations \eqref{Hubble}--\eqref{Raychu} allow one to interpret the new $f(T,L_m)$
contributions as an \emph{effective} modified-gravity (MG) fluid with density
$\rho_{\rm MG}$ and pressure $P_{\rm MG}$. By diffeomorphism invariance, the \emph{total} effective energy--momentum tensor is covariantly conserved, which in FLRW implies the total continuity equation
\begin{equation}
\dot\rho_{\rm tot}+3H\big(\rho_{\rm tot}+P_{\rm tot}\big)=0,
\label{eq:cont_total}
\end{equation}
where $
\rho_{\rm tot}\equiv \rho+\rho_{\rm MG}$ and $P_{\rm tot}\equiv P+P_{\rm MG}.$ It is then convenient to parametrize a possible energy exchange between the
standard component and the MG sector through an interaction term $Q$,
\begin{align}
\dot\rho+3H(\rho+P) &= Q,
\label{eq:cont_matter_Q}\\
\dot\rho_{\rm MG}+3H(\rho_{\rm MG}+P_{\rm MG}) &= -Q,
\label{eq:cont_MG_Q}
\end{align}
so that \eqref{eq:cont_total} is recovered by summing
\eqref{eq:cont_matter_Q} and \eqref{eq:cont_MG_Q}. In GR (or in any theory with
separately conserved matter) one has $Q=0$, while in general $f(T,L_m)$ models
$Q\neq0$ encodes the effective transfer induced by the non-minimal dependence on
$L_m$ (and, when present, the mixed coupling $F_{TL}$).
Using the definitions \eqref{rho_DE}--\eqref{Constrain1}, we define $Q$ explicitly as
\begin{equation}
Q \equiv -\Big[\dot\rho_{\rm MG}+3H(\rho_{\rm MG}+P_{\rm MG})\Big],
\label{eq:Q_def}
\end{equation}
which yields, after exact cancellations and using $T=-6H^2$ (so $\dot T=-12H\dot H$),
the compact implicit identity
\begin{eqnarray}
Q&=&-\frac{1}{16\pi G}\left[F_L\dot\rho +6H F_L(P+L_m)\right. \nonumber\\
&&\left.+(\rho+L_m)F_{LL}\dot L_m -12H\dot H(\rho+L_m)F_{TL} \right].
\label{eq:Q_implicit_general}
\end{eqnarray}

In the radiation epoch we take $w=1/3$ and adopt the matter Lagrangian choice
$L_m=P=\rho/3$, so that $\dot L_m=\dot\rho/3$. In this case the matter balance
equation reads
\begin{equation}
Q_{\rm rad}\equiv \dot\rho+4H\rho,
\label{eq:Qrad_def_short2}
\end{equation}
while the general identity obtained from
$Q\equiv-[\dot\rho_{\rm MG}+3H(\rho_{\rm MG}+P_{\rm MG})]$ reduces to an
\emph{implicit} relation between $Q$, $\dot\rho$ and $\dot H$,
\begin{align}
Q_{\rm rad}&=-\frac{1}{16\pi G}\left[\left(F_L+\frac{4\rho}{9}F_{LL}\right)\dot\rho
+4H\rho\,F_L \right.\nonumber\\
&\left.-16H\rho\,F_{TL}\dot H
\right]\,.
\label{eq:Qrad_implicit_short2}
\end{align}

For separable models ($F_{TL}=0$), Eq.~\eqref{eq:Qrad_implicit_short2} already closes.
However, when $F_{TL}\neq 0$ the Raychaudhuri equation couples $\dot H$ to $\dot\rho$
through the term $H F_{TL}\dot L_m$. By writing the Raychaudhuri equation as
\begin{equation}
\mathcal D\,\dot H
=
-\frac{\rho}{3}\big(16\pi G+F_L\big)
-\frac{H}{3}F_{TL}\dot\rho,
\end{equation}
where
\begin{equation}
\mathcal D\equiv 1+F_T-12H^2F_{TT},
\label{eq:Hdot_rad_short2}
\end{equation}
we can eliminate $\dot H$ from \eqref{eq:Qrad_implicit_short2}. Together with the definition \eqref{eq:Qrad_def_short2}, this yields a linear system for the two unknowns $(\dot\rho,Q_{\rm rad})$. Solving this system and simplifying leads to the closed expression
\begin{equation}
Q_{\rm rad}
=
\frac{\mathcal{A}_{\rm rad}}
{9\mathcal D(16\pi G+F_L)+4\mathcal D\,\rho F_{LL}+48H^2\rho\,F_{TL}^2},
\label{eq:Qrad_master}
\end{equation}
where
\begin{equation}
    \mathcal{A}_{\rm rad}=16H\rho^2\Big(\mathcal D\,F_{LL}+12H^2F_{TL}^2-3F_{TL}(16\pi G+F_L)\Big).
\end{equation}

Equation~\eqref{eq:Qrad_master} is valid for any $f(T,L_m)$ in the radiation era under the choice
$L_m=P=\rho/3$.

{Equations \eqref{eq:Qrad_def_short2}--\eqref{eq:Qrad_master} show that, in general,
the radiation bath does not redshift purely by Hubble dilution. Indeed,
\eqref{eq:Qrad_def_short2} can be written as
\begin{equation}
\frac{d\ln\rho}{d\ln a}=-4+\frac{Q_{\rm rad}}{H\rho},
\label{eq:rho_a_relation}
\end{equation}
so that any nonzero $Q_{\rm rad}$ changes the standard scaling $\rho\propto a^{-4}$.
Equivalently, because $\rho(T)\propto g_*(T)\,T^4$, a sizable $Q_{\rm rad}$ induces
a modified temperature--scale factor relation $T(a)$ (or $t(T)$), i.e.\ the thermal
history is no longer the standard radiation-dominated one. This point has been
emphasized in related non-conserved cosmologies, where the deviation from the
standard $T\propto a^{-1}$ law is explicitly sourced by the interaction term
(see e.g.\ Refs.~\cite{Martins:2025kzd,Haghani:2021fpx}.)}

{This has direct implications for BBN. Standard BBN calculations evolve the weak
rates and nuclear network using $T$ as the time variable and assume the usual
mapping between $T$ and the expansion history (up to the standard $g_*(T)$
variations). If $|Q_{\rm rad}|$ becomes comparable to the Hubble dilution term
$4H\rho$, then the mapping between $T$ and $t$ (or $a$) is modified and one must
solve the coupled non-adiabatic system for $\rho(T)$ (and entropy transfer)
together with the nuclear network. Since the present work follows the standard temperature-based BBN formalism, we restrict our analysis to the regime in which $Q_{\rm rad}$ is a small perturbation to the adiabatic radiation evolution,
quantified by
\begin{equation}
\varepsilon(T)\equiv \left|\frac{Q_{\rm rad}}{4H\rho}\right|\ll 1,
\label{eq:eps_BBN}
\end{equation}
throughout the BBN window. Within this self-consistent adiabatic regime the usual BBN treatment applies, and the abundance constraints can be interpreted as bounds on the EFT parameters of $f(T,L_m)$. }

{\subsection{Late-time limit and connection to torsion-only cosmology}}
\label{subsec:latetime_limit}
{
Although our constraints are derived from the radiation era, it is important to verify that the same $f(T,L_m)$ models admit a consistent late-time background limit. Here we provide a minimal late-time analysis whose purpose is \emph{not} to perform a global fit to late-time data, but rather to show how the explicit $L_m$-dependent sector decouples as radiation dilutes, thereby reducing the theory to the torsion-only sector (TEGR or an $f(T)$ deformation, depending on the benchmark). In this section we assume the standard two-fluid composition (dust plus radiation): 
\begin{equation}
\rho(z)=\rho_m(z)+\rho_r(z),\quad
P(z)=P_m(z)+P_r(z)=\frac{\rho_r(z)}{3},
\end{equation}
with the usual redshift scalings $\rho_m\propto(1+z)^3$ and $\rho_r\propto(1+z)^4$. This assumption is valid in the
adiabatic regime where the radiation interaction term is a small perturbation,
$\varepsilon(z)\equiv \left|Q_{\rm rad}/(4H\rho_r)\right|\ll 1$. We explicitly verify in the BBN analysis
(Sec.~\ref{sec:BBN_FTL}) that this condition holds in the parameter region compatible with the
primordial abundances, so that the standard scalings provide a self-consistent approximation for
both the radiation-era mapping and the late-time limit.}

{For the adopted choice of $L_m=P$ one has
\begin{equation}
L_m(z)=P(z)=\frac{\rho_r(z)}{3}\,,
\label{eq:Lm_twofluid}
\end{equation}
which is sizeable during BBN (radiation domination) but dilutes at late times. It is convenient to introduce the dimensionless expansion function
\begin{equation}
E(z)\equiv \frac{H(z)}{H_0},
\end{equation}
together with the reference scales
\begin{equation}
T_0\equiv -6H_0^2,\qquad L_0\equiv 3M_{\rm Pl}^2H_0^2,
\end{equation}
so that the torsion scalar is $T=-6H^2=T_0\,E^2$.}

{We also define the standard GR dimensionless density combination
\begin{equation}
S(z)\equiv \frac{\rho(z)}{3M_{\rm Pl}^2H_0^2}
=\Omega_{m0}(1+z)^3+\Omega_{r0}(1+z)^4,
\label{eq:S_def}
\end{equation}
and the dimensionless matter-Lagrangian variable
\begin{equation}
\ell(z)\equiv \frac{L_m(z)}{L_0}
=\frac{1}{3}\frac{\rho_r(z)}{3M_{\rm Pl}^2H_0^2}
=\frac{\Omega_{r0}}{3}(1+z)^4.
\label{eq:ell_def}
\end{equation}}

{Note that $\ell(z)\to 0$ as $z\to 0$ (radiation dilution), whereas $\ell(z)$ is large in the BBN window.
Therefore, under $L_m=P$, the explicit $L_m$-dependent corrections are naturally ``pressure-activated'': they are relevant at early times and decouple towards late times.}

Starting from the first modified Friedmann equation, Eq.~\eqref{first Friedmann} and dividing by $H_0^2$, using $\rho=3M_{\rm Pl}^2H_0^2S(z)$, $L_m=L_0\ell(z)$ and $T=T_0E^2$,
we obtain the dimensionless master equation
\begin{equation}
E^2 = S -2E^2 F_T
-\frac{1}{6H_0^2}f
+\frac{1}{6H_0^2}F_L\big(\rho+L_m\big),
\label{eq:dimensionless_master}
\end{equation}
where $F_T\equiv \partial f/\partial T$ and $F_L\equiv \partial f/\partial L_m$ are evaluated on $T=T_0E^2$ and $L_m=L_0\ell$. For each benchmark model we obtain an implicit algebraic equation for $E(z)$, whose late-time limit follows directly from $\ell(z)\to0$.

A potential concern is whether constraining $f(T,L_m)$ models using only BBN is meaningful without performing a global late-time background fit. Here we stress that the aim of this work is precisely to treat $f(T,L_m)$ as an \emph{early-Universe EFT parametrization} and to use BBN as a high-energy probe of departures from the GR expansion rate during the MeV era. In this sense, our results should be interpreted as \emph{necessary early-time viability conditions} on the effective couplings controlling torsion--matter corrections in the radiation epoch, rather than as a complete cosmological model calibrated from recombination to $z\simeq0$. This perspective is standard in BBN-based tests of modified gravity, where one constrains the expansion history in the nucleosynthesis window without requiring a simultaneous SNe/BAO/CMB fit, see e.g.~\cite{Ge:2024tsx,Capozziello:2017bxm,Mishra:2023onl,Asimakis:2022kfk,Anagnostopoulos:2022gej}.

\vspace{+1em}

\section{Big Bang Nucleosynthesis}\label{Standart_BBN}

\subsection{Standard BBN}

 The energy density of relativistic particles and massless radiation is given by
\begin{equation}
    \rho_r=\frac{\pi^2 g_*\mathcal{T}^4}{30}\label{radiation_energy}\,,
\end{equation}
where $g_* \sim 10$ is the effective number of degrees of freedom and $\mathcal{T}$ is the temperature of the state. The Friedmann equations reads
\begin{equation}
    H_{GR}^2=\frac{\rho}{3M_{\rm Pl}^2}\,,
\end{equation}
where $M_{\rm Pl} = (8\pi G)^{-1/2}$ denotes the Planck mass, $H_{\rm GR}$ is the standard Hubble parameter in GR, and $\rho = \rho_m + \rho_r + \rho_\Lambda$, with $\rho_m$ representing the energy density of non-relativistic matter, $\rho_r$ the energy density of radiation, and $\rho_\Lambda$ the energy density associated with dark energy. Since BBN occurs during the radiation-dominated era, $\rho_r$ is the dominant contribution, allowing us to approximate the previous equation as
\begin{equation}
    H_{GR}^2\approx \frac{\rho_r}{3M_{\rm Pl}^2}\label{HGR}\,.
\end{equation}
Using Eq.~\eqref{radiation_energy} we can rewrite $H_{GR}$ as
\begin{equation}\label{H_GR}
    H_{GR}= \sqrt{\frac{\pi^2 g_*}{90}}\frac{\mathcal{T}^2}{M_{\rm Pl}}\,.
\end{equation}

The scale factor evolves as $a\sim t^{1/2}$, with $t$ being the cosmic time, and consequently we can relate the time with temperature using the former equation
\begin{equation}
    \frac{1}{t}\sim \sqrt{\frac{2\pi^2g_*}{45}}\frac{\mathcal{T}^2}{M_{\rm Pl}}\,.
\end{equation}

For temperatures $\mathcal{T} \gtrsim 10\ \mathrm{MeV}$, the energy and number densities of the Universe were dominated by relativistic leptons and photons. During this stage, neutrons and protons remained in thermal equilibrium through weak interactions with the relativistic leptons
\begin{equation}
    \nu_e + n \longleftrightarrow  p +e^-\,,
\end{equation}
\begin{equation}
    e^+ + n \longleftrightarrow  p + \bar{\nu}_e\,,
\end{equation}
\begin{equation}
    n \longleftrightarrow  p + e^-+\bar{\nu}_e\,,
\end{equation}
where $\nu_e$ is the electron neutrino, $n$ is a neutron, $p$ is a proton, $e^-$ is an electron, $e^+$ is a positron and $\bar{\nu}_e$ is the electron antineutrino.

The neutron abundance can be estimated by computing the conversion rate of protons into neutrons $\lambda_{pn}(\mathcal{T})$, and its inverse reaction $\lambda_{np}(\mathcal{T})$. The weak interaction rates  are given by
\begin{equation}
    \Gamma(\mathcal{T})=\lambda_{np}(\mathcal{T})+\lambda_{pn}(\mathcal{T})\,,
\end{equation}
where
\begin{equation}
    \lambda_{pn}(\mathcal{T})=\lambda_{\nu_e + n \to  p +e^-}+\lambda_{e^+ + n \to  p \bar{\nu}_e}+\lambda_{n \to  p + e^-+\bar{\nu}_e}\,,
\end{equation}
and $\lambda_{np}(\mathcal{T})=e^{\mathcal{Q}/\mathcal{T}}\lambda_{pn}(\mathcal{T})$ with $\mathcal{Q}=m_n-m_p=1.29\times10^{-3}\ \mathrm{GeV}$ the mass difference of the neutron and proton. Following the detailed calculation in~\cite{Lambiase:2005kb} , the interaction rate is 
\begin{equation}
    \Gamma(\mathcal{T})=4A\mathcal{T}^3(4!\mathcal{T}^2+2\times 3!\mathcal{Q}\mathcal{T}+2!\mathcal{Q}^2)\,,
\end{equation}
with $A=1.02\times 10^{-11}\ \mathrm{GeV}^{-4}$.

During the freeze-out stage, we can approximate the decoupling temperature $\mathcal{T}_f$ of the various relativistic species as $\mathcal{T}_{\gamma}=\mathcal{T}_{\nu}=\mathcal{T}_{e}=\mathcal{T}_f$. This temperature is reached when the Hubble rate becomes comparable to the weak interaction rate
\begin{equation}\label{qTf}
    H_{GR}(\mathcal{T}_f)\approx\Gamma(\mathcal{T}_f)\,.
\end{equation}

Solving Eq.~\eqref{qTf} for $\mathcal{T}_f$ yields
\begin{equation}
    \mathcal{T}_f\approx 0.55\,{\rm MeV}\,.
\end{equation}

The primordial mass fraction of $^4He$ can be calculated using
\begin{equation}
    Y_p \equiv \lambda\frac{2x(t_f)}{1+x(t_f)}\,,
\end{equation}
where $\lambda=e^{-(t_n-t_f)/\tau}$, with $t_n\approx169\,$s being the freeze-out time of nucleosynthesis, $t_f\approx1.36\,$s the freeze-out time of the weak interactions and $\tau\approx 878.4\,$s the neutron mean lifetime. The function $\lambda(t_f)$ can be interpreted as the fraction of neutrons that decay into protons during the interval $t \in [t_f,t_n]$ and $x(t_f)=e^{-\mathcal{Q}/\mathcal{T}(t_f)}$ is the neutron-to-proton equilibrium ratio. Deviations in $Y_p$ due to variations in the freezing temperature $\mathcal{T}_f$ are given by
\begin{equation}
    \Delta Y_p=Y_p\left[\left(1-\frac{Y_p}{2\lambda}\right)\ln{\left(\frac{2\lambda}{Y_p}-1\right)}-\frac{2t_f}{\tau}\right]\frac{\Delta \mathcal{T}_f}{\mathcal{T}_f}\,.
\end{equation}

According to \cite{Kirkman:2003uv,Izotov:1999wa,Izotov:2003xn,Coc:2004ij,Fields:1998gv,Izotov:1998mj,Olive:1996zu}, the experimental estimations of the value of $Y_p$ for $^4He$ and its uncertainty are respectively  $Y_p=0.2476$ and $|\Delta Y_p|<10^{-4}$. With these values, the variations in temperature can be computed using the previous expression
\begin{equation}
    \left|\frac{\Delta \mathcal{T}_f}{\mathcal{T}_f}\right|< 4.7\times10^{-4}\label{unc}\,.
\end{equation}

\subsection{BBN in modified gravity}\label{BBN}

In modified gravity, the additional contribution $\rho_{\rm MG}$ can be treated as a small perturbation to the standard GR energy density. This perturbation induces deviations in the Hubble parameter, which can in turn be used to compute the resulting shift in the neutron–proton freeze-out temperature, $\Delta \mathcal{T}_f$, thereby providing a means to constrain modified gravity models. Using Eq.~\eqref{Hubble}, the Hubble parameter in the modified theory can be related to its GR counterpart, $H_{\rm GR}$
\begin{equation}
    H=H_{GR}\sqrt{1+\frac{\rho_{MG}}{\rho_r}}\,.
\end{equation}

Expanding the previous expression to first order, the deviation in the Hubble rate, $\Delta H$, is given by 
\begin{equation}\label{Delta_H}
    \Delta H=H-H_{GR}\approx\frac{\rho_{MG}}{\rho_r}\frac{H_{GR}}{2}\,.
\end{equation}

Now, deviations to the interaction rate $\Gamma$ at the freeze-out temperature can be calculated by varying \eqref{qTf}, which provides
\begin{equation}
    \Delta H(\mathcal{T}_f)\approx \Delta \Gamma(\mathcal{T}_f)=24A\mathcal{T}_f^2(20\mathcal{T}_f^2 +8\mathcal{T}_f\mathcal{Q}+\mathcal{Q}^2)\Delta \mathcal{T}_f\label{Gammma_deviation}\,.
\end{equation}

Thus, using Eqs.~\eqref{Delta_H} and \eqref{Gammma_deviation} we find
\begin{equation}
    \frac{\Delta \mathcal{T}_f}{\mathcal{T}_f}=\frac{\rho_{MG}}{\rho_r}\frac{H_{GR}}{48A\mathcal{T}_f^3(20\mathcal{T}_f^2 +8\mathcal{T}_f\mathcal{Q}+\mathcal{Q}^2)}\label{freeze-out_constrain}\,.
\end{equation}

In this work, we study the primordial abundances of deuterium, helium, and lithium within the framework of modified gravity. Thus, to facilitate this analysis, we introduce the following notation:
\begin{equation}
    Z \equiv \frac{H}{H_{GR}}\label{Z}\,.
\end{equation}

Deuterium $^2H$ production in the early Universe occurred through the nuclear reaction~\cite{Dodelson:2003ft}
\begin{equation}
    n+p\to\ ^2H + \gamma\label{Deuterium}\,,
\end{equation}
where $\gamma$ is the photon and $n$ and $p$ are the neutron and the proton, respectively. The numerical best fit for the Deuterium abundance \cite{Steigman:2012ve} is obtained as 
\begin{equation}
    Y_{D_p}=2.6(1\pm 0.06)\left(\frac{6}{\eta_{10}-6(Z-1)}\right)^{1.6}\,,
\end{equation}
with $n_{10}\equiv 10^{10} n_B/n_\gamma\approx 6$, where the ratio $n_B/n_\gamma$ represents the baryon-to-photon ratio \cite{Komatsu_2011}. 

The observational constraint on the Deuterium abundance is $Y_{D_p}=2.55\pm 0.03$ \cite{Fields:2019pfx}. Hence, the constraint on $Z$ is
\begin{equation}
    Z=1.062\pm 0.444\label{D_constrain}\,.
\end{equation}

The production of Helium $^4He$ \cite{Dodelson:2003ft} occurs with the fusion of two Deuterium \eqref{Deuterium} atoms to form the Helium isotope ($^3He$) or Tritium ($^3H$)
\begin{equation}
    ^2H+\ ^2H\to\ ^3He+n\,,
\end{equation}
\begin{equation}
    ^2H+\ ^2H\to\ ^3H+p\,.
\end{equation}
These elements are then fused with $^2H$ to form $^4He$
\begin{equation}
    ^2H+\ ^3H\to\ ^4He+n\,,
\end{equation}
\begin{equation}
    ^2H+\ ^3He\to\ ^4He+p\,.
\end{equation}

The numerical best fit for $^4He$ abundance \cite{Kneller:2004jz, Steigman:2007xt} is obtained as 
\begin{equation}
    Y_p=0.2485\pm 0.0006+0.0016[(\eta_{10}-6)+100(Z-1)]\,.
\end{equation}

The observational constraint on the helium-4 abundance is $Y_{p}=0.2449\pm 0.0040$ \cite{Fields:2019pfx}, which restricts the $Z$ range to
\begin{equation}
    Z=1.0475 \pm 0.105\label{He_constrain}\,.
\end{equation}

For Lithium $^7Li$, the production occurs through the nuclear reactions
\begin{equation}
    ^3H + ^4He\to\ ^7Li+\gamma\,,
\end{equation}
and
\begin{equation}
    ^3He+\ ^4He\to\ ^7Be+\gamma\, ,
\end{equation}
\begin{equation}
    ^7Be+n\to\ ^7Li+p\, .
\end{equation}

The numerical best fit for $^7Li$ abundance \cite{Steigman:2012ve,Steigman:2007xt} is obtained as
\begin{equation}
    Y_{Li}=4.82(1\pm0.1)\left[\frac{\eta_{10}-3(Z-1)}{6}\right]^2\label{57}\,.
\end{equation}

The observational constraint on the lithium abundance is $Y_{Li}=1.6\pm 0.3$ \cite{Fields:2019pfx}. This constrains $Z$ in the range
\begin{equation}
    Z=1.960025 \pm 0.076675\label{Li_constrain}\, .
\end{equation}

The observed value for the lithium abundance is $(2.4-4.3)$ times lower than the predicted value in the standard $\Lambda$CDM BBN. This result is unexpected as the $\eta_{10}$ parameter successfully fits the abundance of $D$ and $^4 He$. Hence, the Lithium constrain for the $Z$ parameter \eqref{Li_constrain} does not overlap with constraints for the other light elements \eqref{D_constrain} \eqref{He_constrain}. This discrepancy is known as the Lithium Problem \cite{Fields:2011zzb}. {Because of this, we will not consider Lithium to constraint the models. Moreover, an analysis intended to address the Lithium Problem is outside the scope of this work.}

\section{BBN constraints on $f(T,L_m)$ gravity}\label{sec:BBN_FTL}

Our methodology proceeds as follows: First, we derive the modified Friedmann equation corresponding to each model under consideration, which encodes the impact of the extra degrees of freedom introduced by the theory. Next, we determine the allowed parameter space by imposing observational bounds, both from deviations in the freeze-out temperature and from the $Z$-constraints associated with the primordial abundances of the light elements studied in this work. It is worth emphasizing that the different BBN constraints used here do not
play identical roles. The deuterium and ${}^4{\rm He}$ bounds on $Z\equiv H/H_{\rm GR}$ provide abundance-level constraints on the modified expansion rate, whereas the freeze-out bound constrains the local shift of the weak decoupling temperature. We therefore impose them as complementary consistency requirements rather than as statistically independent likelihoods. In the benchmark models considered below, the freeze-out condition turns out a posteriori to be the most restrictive constraint on the parameter space. Nevertheless, the deuterium and ${}^4{\rm He}$ bands are retained explicitly because this hierarchy is model- and parameter-dependent and because they verify that the freeze-out-allowed region is also compatible with the observationally inferred light-element abundances. Thus, the final allowed regions should be interpreted as the intersection of necessary BBN consistency conditions, with the freeze-out bound providing the dominant cut for the specific benchmarks studied here.

To make this procedure explicit, we now describe how the cosmological equations are converted into the parameter-space constraints shown below. All BBN bounds are imposed at the characteristic freeze-out temperature $\mathcal{T}_f$. Since the analysis is perturbative around the standard radiation-dominated
solution, the modified-gravity contribution is evaluated on the GR background, $H=H_{\rm GR}(\mathcal{T}_f)$, with $H_{\rm GR}$ given by Eq.~\eqref{H_GR}. The modified Friedmann equation then becomes an algebraic
relation between the free parameters of the theory and the deviation of the
expansion rate from its GR value.

The freeze-out constraint is obtained by using Eq.~\eqref{freeze-out_constrain}
together with the bound in Eq.~\eqref{unc}. The two signs of
$\Delta\mathcal{T}_f/\mathcal{T}_f$ define the upper and lower boundaries of
the freeze-out band in parameter space. The deuterium and ${}^4{\rm He}$
constraints are obtained independently by imposing
\begin{equation}
H^2=Z^2H_{\rm GR}^2
\end{equation}
in the same Friedmann equation, with $Z$ restricted to the observational
intervals in Eqs.~\eqref{D_constrain} and \eqref{He_constrain}. The plotted
allowed region is therefore the intersection of the freeze-out, deuterium and
helium-4 bands, further restricted by the adiabaticity condition
$\varepsilon<0.01$ when applicable.

In many BBN modified-gravity background studies it is convenient to introduce the dimensionless
expansion rate $E(z)\equiv H(z)/H_0$ and impose the present-day normalization condition
$E(0)=1$, which encodes the observational fact $H(z{=}0)=H_0$~\cite{Mishra:2023onl,Anagnostopoulos:2022gej,Sultan:2025tws}. In concrete models, inserting
$z=0$ into the modified Friedmann equation yields an algebraic relation among the model
parameters, allowing one parameter to be eliminated (e.g. $\alpha=\alpha(\beta)$ for fixed
$\Omega_{m0}$ and $\Omega_{r0}$). This procedure is useful when one aims at a global
background fit from early to late times.

{In the present work, however, our primary goal is to constrain EFT-like corrections in the
radiation era using a semi-analytical BBN mapping evaluated at a characteristic temperature
$\mathcal{T}_f$. We therefore keep the EFT couplings independent and treat $E(0)=1$ as an optional
late-time matching condition that can be imposed if one wishes to connect the BBN-allowed
region to late-time cosmology. For completeness, we indicate how this reduction can be
implemented in our parametrization.
}

\subsection{Model I: separable torsion--matter correction}

{For the first model we consider
\begin{equation}
f(T,L_m)=T_0\,\alpha\left(\frac{T}{T_0}\right)^{n}
\;+\;
T_0\,\beta\left(\frac{L_m}{L_0}\right)^{1+\eta}\,,
\label{eq:modelI_intro}
\end{equation}}

{The first term is a standard and widely studied $f(T)$-type deformation of TEGR, often employed in context of inflation, to mimic an effective dark-energy component and to model late-time acceleration through a torsion-driven modification of the background dynamics (see the review \cite{Li:2018ixg} and references therein). The second term is an EFT-inspired matter-sector correction: choosing $1+\eta$ where for $\eta$ close to unity provides the minimal deviation from minimal matter coupling while keeping the modification controlled and for higher values more robust high-energy modifications. This structure is useful both at early times (where high densities can enhance the sensitivity to the $\beta$-term) and at late times (where the $\alpha$-term can dominate and drive acceleration), thereby offering a minimal two-parameter template that bridges early- and late-Universe phenomenology within a single $f(T,L_m)$ framework. Moreover, it mirrors the logic of curvature-based extensions in which a ``geometrical'' sector is supplemented by non-minimal matter operators, as in the
$f(R,L_m)$ literature \cite{Bertolami:2008zh}.}

{\subsubsection{Late-time limit}}
{Using $T=T_0E^2$ and $\ell=L_m/L_0$, the model in question can be written as
\begin{equation}
f(T,L_m)=T_0\Big[\alpha (E^2)^n+\beta\,\ell^{\,1+\eta}\Big]\,.
\end{equation}
that by substituting into Eq.~\eqref{eq:dimensionless_master} and using $-T_0/(6H_0^2)=1$ yields
\begin{equation}
E^2=S+\alpha(1-2n)(E^2)^n+\beta\left[\ell^{1+\eta}-(1+\eta)\ell^{\eta}(S+\ell)\right]\,,
\label{eq:modelI_E_equation}
\end{equation}}

{Since $\ell\to0$, for $\eta>0$ the explicit $L_m$-dependent piece vanishes and the model reduces to a torsion-only $f(T)$ background equation,
\begin{equation}\label{eq:modelI_latetime}
E^2=S+\alpha(1-2n)(E^2)^n
\qquad (\ell\to0,\ \eta>0).
\end{equation}}

{\subsubsection{BBN constraints}}
Inserting the model~\eqref{eq:modelI_intro} into Eq.~\eqref{first Friedmann} we obtain the first modified Friedemann equation
\begin{equation}
\begin{split}
 H^2
&=
\frac{\rho}{3M_{\rm Pl}^2}
+\alpha T_0
\left(\frac{-6H^2}{T_0}\right)^n
\left(\frac{2n-1}{6}\right)
+\\
&
\frac{T_0\beta}{6}
\left(\frac{\rho}{3L_0}\right)^{\eta}
\left[
\frac{4}{3}(1+\eta)\frac{\rho}{L_0}
-
\frac{\rho}{3L_0}
\right].
\end{split}\label{first_model_friedmann}
\end{equation}
Making use of Eqs.\eqref{first_model_friedmann} and \eqref{freeze-out_constrain} the parameter $\beta$ can be expressed in term of $\alpha$
\begin{equation}
\beta=\frac{\mathcal N_{\rm fo}}{\mathcal D_{\rm fo}},
\label{210_freeze-out}
\end{equation}
where
\begin{equation}
\begin{aligned}
\mathcal N_{\rm fo}
&=
\frac{\rho\,48A\mathcal{T}_f^3
\left(20\mathcal{T}_f^2+8\mathcal{T}_f\mathcal{Q}+\mathcal{Q}^2\right)}
{3 M_{\rm Pl}^2\,H_{\rm GR}}
\frac{\Delta \mathcal{T}_f}{\mathcal{T}_f}
\\
&
-\frac{\alpha T_0}{6}
\left(\frac{-6H_{\rm GR}^2}{T_0}\right)^n(2n-1),
\end{aligned}
\end{equation}
and
\begin{equation}
\mathcal D_{\rm fo}
=
\frac{T_0}{6}
\left(\frac{\rho}{3L_0}\right)^{\eta}
\left[
\frac{4}{3}(1+\eta)\frac{\rho}{L_0}
-\frac{\rho}{3L_0}
\right].
\end{equation}

\begin{figure}[t!]
    \centering
\includegraphics[width=\columnwidth]{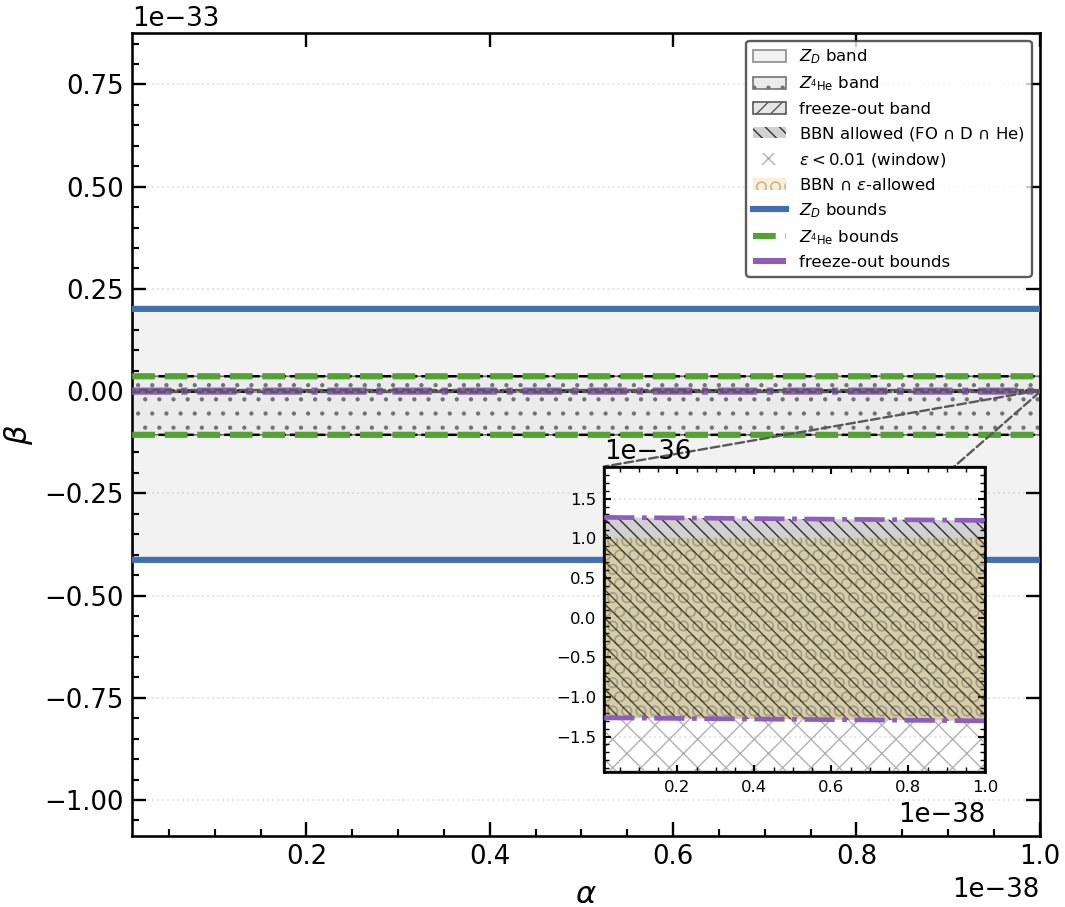}
    \caption{Combined freeze-out, deuterium and helium-4 $Z$ constraints for the free parameters $(\alpha,\beta)$, with $n=2$ and $\eta=1$.}
    \label{fig:210_Zbounds}
\end{figure}

Similarly, we can use \eqref{Z} to write another constraint for $\beta$
\begin{equation}
    \beta=\frac{Z^2\,H_{\mathrm{GR}}^2\;-\;\frac{M_{\mathrm{pl}}^{-2}}{3}\,\rho\;-\;\frac{\alpha T_0}{6}\left(\frac{-6 H_{\mathrm{GR}}^2}{T_0}\right)^{n}(2n-1)}{\mathcal D_{\rm fo}}\,, \label{210_beta_Z}
\end{equation}
with $Z$ being the constraint associated with a specific light element.
Figure~\ref{fig:210_Zbounds} displays the freeze-out, deuterium and ${}^4{\rm He}$ constraints in the $(\alpha,\beta)$ plane for Model~I. For the benchmark choice $n=2$ and $\eta=1$, the freeze-out condition gives the narrowest band and therefore dominates the final allowed region. The deuterium and helium-4 constraints are nevertheless shown because they provide an independent abundance-level check on the expansion rate and confirm that the freeze-out-compatible region also lies within the observationally allowed BBN range. The shaded region denotes the intersection of these consistency requirements, further restricted by the adiabaticity condition $\varepsilon<0.01$. The inset shows a magnified view of the physically relevant region, illustrating that the viable parameter space is extremely narrow. Numerically, we find $\beta\in [-1.261,1]\times 10^{-36}$ and $\alpha_{\rm max}=1.07896\times 10^{-36}$ under the combined BBN and adiabaticity requirements.

{For the late-time limit~\eqref{eq:modelI_latetime} the benchmark used in this work, $n=2$, gives 
\begin{equation}
E^2 \;=\; S - 3\alpha (E^2)^2,
\qquad\Longleftrightarrow\qquad
3\alpha (E^2)^2 + E^2 - S = 0.
\label{eq:modelI_quadratic}
\end{equation}
that can be solved $E^2$ and selecting the branch continuously connected to GR (the one that remains finite as $\alpha\to 0$), giving
\begin{equation}
E^2
=
\frac{-1+\sqrt{\,1+12\alpha S\,}}{6\alpha}.
\label{eq:E2_exact}
\end{equation}}

{It is convenient to rewrite \eqref{eq:E2_exact} in a form that allows a clean control of the deviation
from GR without any assumptions on the magnitudes of $E$ or $S$. Using the identity
$\sqrt{1+u}-1=u/(1+\sqrt{1+u})$ with $u\equiv 12\alpha S$, one finds
\begin{equation}
E^2 = \frac{2S}{1+\sqrt{1+u}},
\label{eq:E2_exact_rewrite}
\end{equation}
and therefore the fractional deviation from the GR expression $E^2=S$ is
\begin{equation}
\frac{E^2-S}{S} = \frac{2}{1+\sqrt{1+u}}-1 = -\frac{u}{\big(1+\sqrt{1+u}\big)^2}.
\label{eq:frac_dev_exact}
\end{equation}}

{Since $\sqrt{1+u}\ge 1$ for $u\ge 0$, we have $\big(1+\sqrt{1+u}\big)^2\ge 4$, and thus we obtain the bound
\begin{equation}
\left|\frac{E^2-S}{S}\right| = \frac{u}{\big(1+\sqrt{1+u}\big)^2} \le \frac{u}{4} =3\alpha S,
\label{eq:frac_dev_bound}
\end{equation}
that is fully non-perturbative and makes no assumption on the size of $E$ or $S$: the departure from the GR background is controlled directly by the single combination $\alpha S(z)$.
In particular, for the obtained value $\alpha_{\text{max}}$ one obtains
\begin{equation}
\left|\frac{E^2-S}{S}\right| \le 3\alpha_{\text{max}} S(z) = 3.23688\times 10^{-36}\, S(z),
\end{equation}
which is negligibly small throughout the late-time cosmological evolution (and remains tiny even for very large $S$, e.g.\ deep into the radiation era), demonstrating that the background expansion effectively coincides with the GR one on the physical branch.}

\subsection{Model II: mixed torsion--matter monomial}

{To probe genuinely non-separable torsion--matter interactions we adopt the monomial
\begin{equation}
f(T,L_m)
=
\lambda T_0
\left(\frac{T}{T_0}\right)^p
\left(\frac{L_m}{L_0}\right)^q .
\label{eq:modelII_intro}
\end{equation}
where $\lambda$ is dimensionless and $(p,q)$ control the scaling with torsion and matter.
This form is directly motivated by the EFT viewpoint: it represents a leading mixed operator built from the available low-energy scalars once non-minimal couplings are allowed. Mixed operators of this type are the torsional analogue of curvature--matter couplings widely studied in the $f(R,L_m)$ framework \cite{Bertolami:2008zh}, where non-minimality generically induces an effective (non-)conservation of the standard component and can be interpreted as an energy exchange with an effective geometrical sector. In $f(T,L_m)$ gravity the same qualitative feature arises, but with a torsion-based origin; this makes Model~\eqref{eq:modelII_intro} particularly well suited to test the distinctive signature of $f(T,L_m)$ theories in the radiation era, including the appearance of an interaction term $Q$ in the effective continuity equations. Phenomenologically, $(p,q)$ can be tuned
to enhance sensitivity at early times (BBN-scale torsion and density) while remaining subdominant at late times, or vice versa, making this a flexible yet still minimal two-parameter benchmark. This type of non-minimal couplings can aid in the understanding of the asymmetry between matter and anti-matter~\cite{CRUZ2026117304}.}

\subsubsection{Late-time limit}

{With $T=T_0E^2$ and $L_m=L_0\ell$, one finds
\begin{equation}
f(T,L_m)=\lambda\,T_0\,(E^2)^p\,\ell^q,
\end{equation}
that by substituting into Eq.~\eqref{eq:dimensionless_master} gives
\begin{equation}
E^2 = S +\lambda(1-2p)(E^2)^p\ell^q -q\lambda (E^2)^p\ell^{q-1}(S+\ell).
\label{eq:modelII_E_equation}
\end{equation}}

Since $\ell\to0$, the mixed torsion--matter correction decouples whenever $q>1$ (because then $\ell^{q-1}\to0$), and the standard GR background is recovered:
\begin{equation}
E^2\to S
\qquad (\ell\to0,\ q>1).
\label{eq:modelII_latetime_qgt1}
\end{equation}
\subsubsection{BBN constraints}
{Inserting the model~\eqref{eq:modelII_intro} into Eq.~\eqref{first Friedmann} we obtain the first modified Friedemann equation
\begin{equation}
H^2=\frac{\rho}{3M_{\rm Pl}^2}+\frac{\lambda T_0}{6}\left(2p+4q-1\right)\left(\frac{-6H^2}{T_0}\right)^p\left(\frac{\rho}{3L_0}\right)^q .
\label{mixed_friedmann}
\end{equation}
\begin{figure}[t!]
    \centering
    \includegraphics[width=\columnwidth]{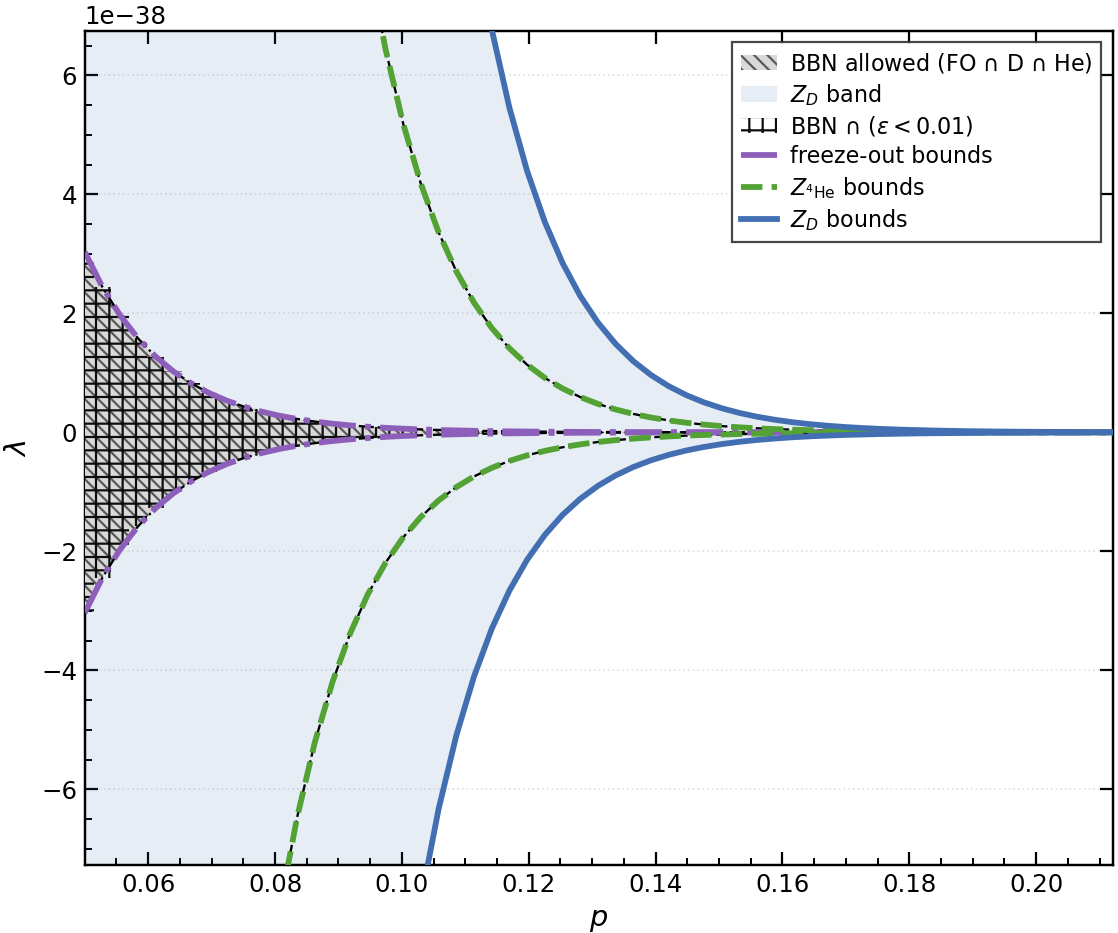}
    \caption{Freeze-out, deuterium and helium-4 $Z$ constraints for Model~II in the $(p,\lambda)$ plane, with $q=2$.}
    \label{fig:mixed}
\end{figure}

Using Eqs.~\eqref{mixed_friedmann} and \eqref{freeze-out_constrain}, the parameter $\lambda$ can be expressed in terms of $p$
\begin{equation}
\lambda
=
\frac{
6\,\Delta_{\rm fo}
}{
(2p+4q-1)T_0
\left(\frac{-6H_{\rm GR}^2}{T_0}\right)^p
\left(\frac{\rho}{3L_0}\right)^q
},
\label{mixed_freeze}
\end{equation}
where we define
\begin{equation}
\Delta_{\rm fo}=\frac{\rho\,48A\mathcal{T}_f^3\left(20\mathcal{T}_f^2+8\mathcal{T}_f\mathcal{Q}+\mathcal{Q}^2\right)}{3M_{\rm Pl}^2H_{\rm GR}}\frac{\Delta\mathcal{T}_f}{\mathcal{T}_f}.
\end{equation}

Similarly, using \eqref{Z} one obtains 
\begin{equation}
\lambda=\frac{6\left[Z^2H_{\rm GR}^2-\frac{\rho}{3M_{\rm Pl}^2}
\right]}{(2p+4q-1)T_0\left(\frac{-6H_{\rm GR}^2}{T_0}\right)^p\left(\frac{\rho}{3L_0}\right)^q}.\label{mixed_Z}
\end{equation}

Fig.~\ref{fig:mixed} displays the freeze-out, deuterium and ${}^4{\rm He}$ constraints in the $(p,\lambda)$ plane for Model~II, with $q=2$. As in Model~I, the freeze-out condition gives the narrowest band over most of the displayed range and therefore provides the dominant restriction on the final allowed region. The deuterium and helium-4 bands are broader, but they are retained as independent abundance-level consistency checks on the modified expansion rate. The shaded region corresponds to the intersection
of the BBN consistency requirements.

The allowed interval for $\lambda$ decreases rapidly as $p$ increases, approaching zero for larger values of $p$. This behavior reflects the strong suppression of higher-order torsional corrections in the BBN regime: for larger $p$, a smaller value of the coupling $\lambda$ is required in order to keep the modified contribution to $H$ compatible with the freeze-out and
light-element abundance constraints.

\subsection{Model III: logarithmic torsion with screened matter correction}

Finally, we consider a model that combines a simple logarithmic torsional deformation with a matter operator that is dynamically suppressed at large $L_m$:
\begin{align}
f(T,L_m)&=T_0\,\alpha\,\ln\!\left(\frac{T}{T_0}\right)
\;\\
&+\;T_0\,\beta\left(\frac{L_m}{L_0}\right)^{2}
\exp\!\left[-\left(\frac{L_m}{L_0}\right)^{m}\right],
\  m>0\,.
\label{eq:modelIII_intro}
\end{align}

{Logarithmic corrections are a well-motivated and frequently explored functional form in modified gravity, where they can capture ``soft'' departures from the baseline theory and can effectively generate a near-constant contribution over a range of scales; recent analyses in torsional contexts have also examined logarithmic $f(T)$ cosmologies \cite{Aquino:2025sdb}. The matter-sector term in \eqref{eq:modelIII_intro} is designed to implement a simple \emph{screening} (or decoupling) behavior at high densities: for $L_m\gg L_0$ the exponential factor suppresses the correction, while for intermediate densities the correction can become appreciable. From an EFT perspective, such a structure can be viewed as a phenomenological resummation capturing the idea that certain effective operators may be dynamically suppressed in extreme regimes, while remaining relevant in the intermediate-energy window most constrained by cosmological observations. This model is therefore interesting both for early-time applications---where BBN can bound the parameter region in which the screening is not yet complete---and for late-time phenomenology, where the torsional logarithmic term can contribute to the acceleration sector while the matter correction remains controlled.}

\subsubsection{Late-time limit}
Using once more $T=T_0E^2$ and $\ell=L_m/L_0$, we have
\begin{equation}
f(T,L_m)=T_0\alpha\ln(E^2)+T_0\beta\,\ell^2 e^{-\ell^m}.
\end{equation}
that by substituting it into Eq.~\eqref{eq:dimensionless_master} yields the implicit equation
\begin{align}
E^2&=S-2\alpha+\alpha\ln(E^2)-\beta\,e^{-\ell^m}(2\ell-m\ell^{m+1})(S+\ell)\nonumber\\
&+\beta\ell^2e^{-\ell^m}.
\label{eq:modelIII_E_equation}
\end{align}

{Since $\ell\to0$, the explicit $L_m$-dependent terms vanish at least linearly, and the model reduces to a torsion-only logarithmic deformation,
\begin{equation}\label{eq:modelIII_latetime}
E^2 = S -2\alpha +\alpha\ln(E^2) +\mathcal{O}(\ell)
\qquad (\ell\to0).
\end{equation}}

\subsubsection{BBN constraints}

Inserting the model~\eqref{eq:modelIII_intro} into Eq.~\eqref{first Friedmann} we obtain the first modified Friedemann equation
\begin{align}
H^2 &= \frac{M_{\rm Pl}^{-2}}{3}\rho
+ \frac{\alpha T_0}{3}\left(
1-\frac{1}{2}\ln\left(\frac{-6H^2}{T_0}\right)\right)
+\nonumber\\
&\frac{T_0}{6}\beta e^{-\left(\frac{L_m}{L_0}\right)^m}
\left\{ \frac{4\rho}{3L_0}\left(\frac{L_m}{L_0}\right)\left[2 - m \left(\frac{L_m}{L_0}\right)^m\right] \right.\nonumber\\ 
&\left.- \left(\frac{L_m}{L_0}\right)^2 \right\}\label{third_friedamn}
\end{align}

Using of Eqs.\eqref{third_friedamn} and \eqref{freeze-out_constrain} the parameter $\beta$ can be expressed in term of $\alpha$ as
\begin{equation}
    \beta = \frac{\mathcal{N}}{\mathcal{K}}\,,
\end{equation}
where
\begin{align}
\mathcal{N}
&=
\frac{\rho\,48 A
\left(
20\mathcal{T}_f^5
+8\mathcal{T}_f^4\mathcal{Q}
+\mathcal{T}_f^3\mathcal{Q}^2
\right)}
{3 M_{\rm Pl}^2 H_{\rm GR}}
\frac{\Delta \mathcal{T}_f}{\mathcal{T}_f}
\nonumber\\
&\quad
-\frac{T_0\alpha}{3}
\left[
1-\frac{1}{2}
\ln\left(\frac{-6H_{\rm GR}^2}{T_0}\right)
\right],
\end{align}
and
\begin{equation}
    \mathcal{K}= \frac{T_0e^{-\left(\frac{\rho}{3L_0}\right)^m}\left[ \frac{4\rho^2}{9L_0^2}  \left( 2 - m \left(\frac{\rho}{3L_0}\right)^m \right) - \left( \frac{ \rho}{3L_0} \right)^2 \right]}{6}.
\end{equation}

Similarly, using \eqref{Z}
\begin{equation}
    \beta=\frac{Z^2\,H_{\rm GR}^2-\frac{M_{\rm Pl}^{-2}}{3}\rho-\frac{T_0\alpha}{3}\left(1-\frac{1}{2}\ln\left(\frac{-6H_{\rm GR}^2}{T_0}\right)\right)}{\mathcal{K}}
\end{equation}

\begin{figure}[t!]
    \centering
    \includegraphics[width=\columnwidth]{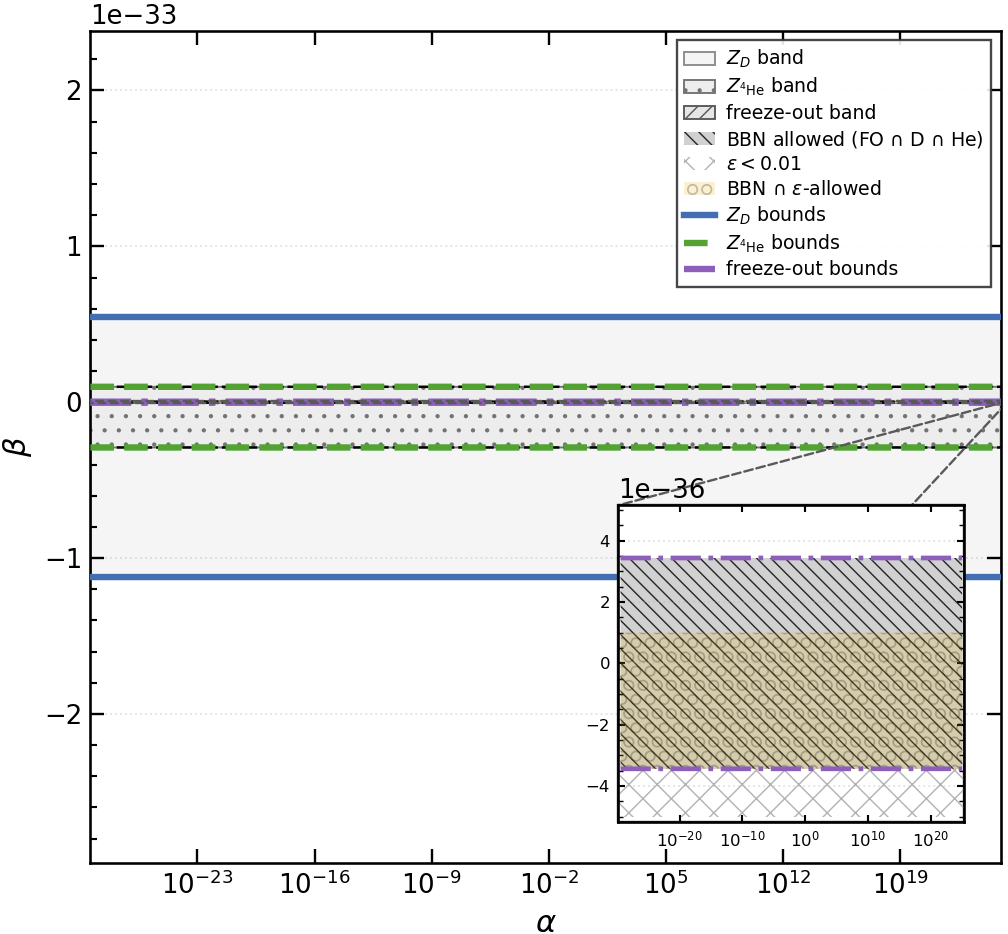}
    \caption{Freeze-out, deuterium and helium-4 $Z$ constraints for Model~III in the $(\alpha,\beta)$ plane, with $m=10^{-7}$.}
    \label{fig:214Z}
\end{figure}

Fig.~\ref{fig:214Z} displays the freeze-out, deuterium and ${}^4{\rm He}$ constraints in the $(\alpha,\beta)$ plane for Model~III, with $m=10^{-7}$. In the displayed region, the freeze-out condition again provides the dominant restriction, while the deuterium and helium-4 bands act as independent abundance-level checks on the modified expansion rate. The freeze-out-compatible region lies within the observationally allowed BBN range selected by the light-element constraints.

Compared with Models~I and II, the logarithmic torsional sector allows a wider range of $\alpha$. Numerically, we find an allowed interval extending up to $\alpha_{\rm max}=2.59259\times 10^{26}$ under the combined BBN requirements. In contrast, the matter-sector coupling remains strongly constrained, with $\beta\in[-3.5,1]\times10^{-36}$. This indicates that, even in the screened logarithmic benchmark, the explicit $L_m$-dependent correction must remain highly suppressed during BBN. The result is consistent with the interpretation that the exponential factor controls the high-density matter correction, while the torsional logarithmic sector can accommodate a larger effective parameter range.

For the late-time limit, using concavity of $\ln$, for any $S>0$ and any $E^2>0$ one has the tangent bounds
\begin{equation}
\ln(E^2)\le \ln S+\frac{E^2-S}{S},
\qquad
\ln(E^2)\ge \ln S+\frac{E^2-S}{E^2}.
\label{eq:log_tangent_bounds_E}
\end{equation}

Inserting the first inequality of \eqref{eq:log_tangent_bounds_E} into
\eqref{eq:modelIII_latetime} and rearranging (for $S>\alpha$) gives
\begin{equation}
E^2 \le \frac{S-3\alpha+\alpha\ln S}{1-\alpha/S}.
\label{eq:E2_upper_concise}
\end{equation}
and by using the second inequality of \eqref{eq:log_tangent_bounds_E} in \eqref{eq:modelIII_latetime} yields
\begin{equation}
(E^2)^2-A\,E^2+\alpha S\ge 0,
\qquad
A\equiv S-\alpha+\alpha\ln S\,,
\label{eq:E2_quadratic}
\end{equation}

If $A^2-4\alpha S\ge 0$, the solutions satisfy $E^2\le \frac{A-\sqrt{A^2-4\alpha S}}{2}$ or
$E^2\ge \frac{A+\sqrt{A^2-4\alpha S}}{2}$. Requiring continuity to GR as $\alpha\to 0$
selects the large-root (GR-connected) branch, hence
\begin{equation}
E^2 \ge \frac{A+\sqrt{A^2-4\alpha S}}{2}.
\label{eq:E2_lower_concise}
\end{equation}

Combining \eqref{eq:E2_upper_concise} and \eqref{eq:E2_lower_concise}, we obtain an explicit bracket for the GR-connected late-time solution,
\begin{equation}
\frac{A+\sqrt{A^2-4\alpha S}}{2}
\;\le\;
E^2
\;\le\;
\frac{S-3\alpha+\alpha\ln S}{1-\alpha/S},
\label{eq:E2_bracket_concise}
\end{equation}
valid for $S>\alpha$ and $A^2-4\alpha S\ge 0$. In particular, for $|\alpha|\ll S$ one has
\begin{equation}
E^2 = S+\alpha(\ln S-2)+\mathcal O\!\left(\frac{\alpha^2}{S}\right),
\label{eq:E2_asymptotic_concise}
\end{equation}
so the deviation from the GR dust+radiation reference is parametrically controlled by $|\alpha|/S$ and is negligible for a considerable range of the interval allowed by our BBN bounds. For higher values of $\alpha$ the theory deviates from GR being of the $f(T)$ family.

\section{Conclusion}\label{Conclusion}
In this paper, we examined the implications of $f(T,{L}_m)$ gravity models for BBN. Focusing on three EFT inspired representative $f(T,{L}_m)$ models, we derived constraints on their free parameters using observational $Z$-bounds from the primordial abundances of deuterium and helium-4, as well as deviations in the neutron–proton freeze-out temperature. For each model, the parameter space was determined to ensure consistency with the observational $Z$-constraints of deuterium, helium-4. These individual constraints were subsequently combined, with the overlapping regions defining the viable parameter space, thereby assessing the overall compatibility of each model with BBN observations. To make the parameter interpretation uniform and to facilitate comparison with the literature, we adopted dimensionless normalizations based on $T_0=-6H_0^2$ and a fixed reference density $L_0$, and we expressed our benchmark models in terms of $T/T_0$ and $L_m/L_0$.

Using a semi-analytical BBN strategy (in the spirit of Ref.~\cite{Ge:2024tsx}), we translated departures from GR into bounds on the modified expansion history through the ratio $Z\equiv H/H_{\rm GR}$ evaluated at a characteristic freeze-out temperature, together with constraints from the neutron--proton freeze-out condition. Our primary constraints were obtained from deuterium and ${}^4$He, which are generally regarded as the most robust probes of the BBN expansion rate. Within this framework we derived, for each model, the corresponding allowed regions in the relevant parameter plane and assessed the overall compatibility of the models with BBN observations. For the benchmark choices studied here, the freeze-out condition provides the dominant restriction on the parameter space, while the deuterium and ${}^4{\rm He}$ bounds act as independent abundance-level consistency checks on the modified expansion rate.

A distinctive feature of $f(T,L_m)$ cosmology is that the explicit $L_m$ dependence can induce an effective energy exchange between the standard component and the modified-gravity sector. To make this effect explicit and to ensure the internal consistency of the radiation-era interpretation underlying the semi-analytical mapping, we derived the general interaction term $Q$ from the effective-fluid decomposition and obtained a closed expression for $Q_{\rm rad}$ in the radiation epoch (including the case with mixed coupling $F_{TL}\neq 0$). We then used the dimensionless ratio $\varepsilon \equiv \left|\frac{Q_{\rm rad}}{4H\rho}\right|$ as a robustness criterion: in the parameter regions compatible with the BBN bounds we find $\varepsilon\ll 1$ (typically at the $10^{-3}$--$10^{-2}$ level, and at most at the few-percent level for our screened logarithmic benchmark), supporting the use of the standard radiation-dominated scaling as an accurate approximation within the allowed regions.

As illustrative EFT-motivated benchmarks we constrained three representative classes of $f(T,L_m)$ functions: (i) a separable model combining a torsion-sector deformation with a near-linear matter correction of the form $f(T,L_m)=T_0\,\alpha (T/T_0)^n + T_0\,\beta (L_m/L_0)^{1+\eta}$; (ii) a genuinely mixed monomial operator, $f(T,L_m)=\lambda T_0 (T/T_0)^p(L_m/L_0)^q$; and (iii) a logarithmic torsional deformation supplemented by an exponentially screened matter correction given by $f(T,L_m)=T_0\,\alpha\ln(T/T_0)+T_0\,\beta(L_m/L_0)^2\exp[-(L_m/L_0)^m]$. For each case we provided the corresponding BBN-allowed parameter regions and the late-time limit. We highlighted how the scaling with torsion and matter controls the sensitivity of BBN to the EFT coefficients. Overall, the existence of viable regions demonstrates that torsion--matter EFT corrections can be compatible with the stringent early-Universe constraints from light-element abundances while remaining self-consistent with a radiation-dominated interpretation.

Finally, we emphasize that the present constraints are intended as transparent first-pass bounds and a useful model-selection guide. A fully comprehensive test would embed the modified background evolution into a complete BBN reaction network and perform a likelihood-based analysis, as implemented in public codes such as \texttt{PArthENoPE}~\cite{Gariazzo_2022} and \texttt{AlterBBN}~\cite{Arbey:2011nf}, including nuclear-rate uncertainties and Bayesian inference as done in~\cite{Sobotka:2023bzr,An:2023buh}. In $f(T,L_m)$ models such an implementation would require, in addition to $H(T)$, a careful treatment of the temperature--scale-factor relation whenever $Q\neq 0$. We leave this full numerical and Bayesian analysis for future work, and we expect that the EFT-motivated templates and parameter priors identified here provide a natural starting point for such extensions and for further confronting torsion--matter couplings with early-Universe data.

\section*{Acknowledgments}
DSP, FSNL and JPM acknowledge support from the Funda\c{c}\~{a}o para a Ci\^{e}ncia e a Tecnologia (FCT) research grants UIDB/04434/2020, UIDP/04434/2020 and PTDC/FIS-AST/0054/2021. FSNL also acknowledges support from the FCT Scientific Employment Stimulus contract with reference CEECINST/00032/2018.

\bibliographystyle{spphys}
\bibliography{biblio}

\end{document}